\documentclass[english,12pt]{article}
\usepackage[cp1251]{inputenc}
\usepackage{babel}
\usepackage{amsfonts, amsmath}

\newcommand{\be}{\begin{equation}} \newcommand{\ee}{\end{equation}}

\begin{document}
\begin{center}
{\bf Probable Entropic Nature of Gravity in Ultraviolet and Infrared Limits—Part I: An Ultraviolet Case}\\
\vspace{5mm} A.E.Shalyt-Margolin \footnote{E-mail:
a.shalyt@mail.ru; alexm@hep.by}\\ \vspace{5mm} \textit{National
Centre of Particles and High Energy Physics, Bogdanovich Str. 153,
Minsk 220040, Belarus}
\end{center}
 \rm\normalsize \vspace{0.5cm}
\begin{abstract}
This work presents a study of the possibility for extending the
well-known results of E.Verlinde concerning the entropic nature of
gravity to the ultraviolet region (Planck’s energies) and also the
derivation of quantum corrections to Einstein Equations.
\end{abstract}

\section{Introduction}
In the last   15-20 years new very interesting approaches to
gravity studies have been proposed, which may be divided into
“thermodynamical” and “theoretical-informational” approaches. The
approach suggested in the pioneer work by T. Jacobson \cite{Jac1}
has been considerably extended in a series of remarkable papers by
T.Padmanabhan \cite{Padm3}--\cite{Padman2009}. The paper by
E.Verlinde \cite{Verlinde} stating a secondary character of the
gravitational interaction and its entropic nature was published in
2010 after the appearance in the ArXiv.
 The paper \cite{Verlinde} introducing such specific terms as "Entropic Force"
 has been followed by numerous studies (e.g. \cite{Verlinde2}--\cite{Verlinde2.10} and
 others).
 \\ In this work the author studies the possibility for extension of
the results given in \cite{Verlinde} to the ultraviolet region (Planck’s energies)
 and presents the derivation of quantum corrections to Einstein Equations using
the dimensionless small parameter $\alpha$ introduced by the
author in his previous works \cite{shalyt1} --
\cite{shalyt-entropy2}.

\section{Fundamental Quantities and Their High-Energy Deformation}
In this section the author uses the  “ideology”, terms, and
notation introduced in \cite{Verlinde} to extend the corresponding
results to the ultraviolet and infrared gravity regions. It may be
stated that the results in \cite{Verlinde}
 have been obtained for the “medium” energies, i.e. for the range of well-known energies,
where the General Relativity (GR) is valid. But owing to modern
knowledge, in the ultraviolet and infrared limits gravity may be
modified. As regards the ultraviolet (Planck) scale, this idea has
been proposed long ago \cite{Gar1} --\cite{Ma1} and in this
situation the word “may” we have to replace by the word “must”. As
for low energies, there are many recent publications considering
the infrared (for great distances) modification of gravity
(e.g.\cite{Rub},\cite{Rub1}). And the modification can have a
solid experimental status in the nearest future \cite{Turysh1}.
\\Naturally, when we are concerned with extension of some results to higher or lower energies,
 the principle of conformity must be executed undeviatingly:  on going
 to the known energy scales the known results must be reproduced.
 \\ In this Section it is shown that the fundamental quantities $A$,$N$,$T$,
 defined in \cite{Verlinde} and associated with the holographic screen (where $A$ - its surface area,
  $N$ - number of data bits "existing"   on $A$, $T$
 -- its temperature), may be supplementary defined for the region
  of high energies so that  at  normal energies they be coincident with  values given  in
 \cite{Verlinde}. In the process we take
the corresponding quantities for the stationary Schwarzshild black
hole as most natural holographic object.
\\The idea is as follows: formulae for the correction of the fundamental quantities
within the Generalized Uncertainty Principle (GUP)
 for the holographic screen $\cal S$ from \cite{Verlinde} are similar to those for the black hole.
\\Let   us  consider  Section  5.2  (Derivation of Einstein Equations) in   \cite{Verlinde}.
 In this Section formula (5.32) for a "bit density" on the holographic screen is given as
\begin{equation}\label{Verl1}
 dN= {dA\over G\hbar}.
\end{equation}
 However, when the holographic principle
\cite{Hooft1}--\cite{Bou3} is valid,
 $N$ is actually the entropy $S$ up to the factor $S\sim N$ and hence from (\ref{Verl1}) it follows directly that
\begin{equation}\label{Verl2}
dS \sim {dA\over G\hbar}.
\end{equation}
What are the changes in $S$ on going to high (Planck) energies?
The answer to this question is already known owing to the fact
 that at these energies the Heisenberg Uncertainty Principle (HUP) is replaced by GUP
\cite{Ven1}--\cite{Kempf}:
\begin{equation}\label{GUP1}
\triangle x\geq\frac{\hbar}{\triangle p}+ \ell^2 \frac{\triangle
p}{\hbar},
\end{equation}
where  $ \ell^2 =\alpha^{\prime} l_{p}^2$ and $\alpha^{\prime}$ is
 the dimensionless numerical factor. The well-known Bekenstein-Hawking
formula for the  black hole entropy  in the semiclassical
approximation \cite{Bek1},\cite{Hawk1}
\begin{equation}\label{GUP4}
S^{BH}=\frac{A}{4l^{2}_{p}}
\end{equation}
is modified by the corresponding quantum corrections on going from
HUP to GUP \cite{Park}--\cite{Nou}.
\\In particular, \cite{mv}:
\begin{equation}\label{GUP5.0}
S^{BH}_{GUP} =\frac{A}{4l_{p}^{2}}-{\pi\alpha^{\prime 2}\over
4}\ln \left(\frac{A}{4l_{p}^{2}}\right) +\sum_{n=1}^{\infty}c_{n}
\left({A\over 4 l_p^2} \right)^{-n}+ \rm{const}\;,
\end{equation}
where the expansion coefficients $c_n\propto \alpha^{\prime
2(n+1)}$ can always be computed to any desired order of accuracy.
\\ The general form of quantum corrections for the black hole entropy derived in (\ref{GUP5.0})
remains valid for any horizon spaces and, in particular, for the
holographic screen  $\cal S$ from \cite{Verlinde}. Specifically,
in \cite{Zhang},\cite{Cai-new} the logarithmic correction was
obtained in the following form:
\begin{equation}\label{GUP5.1}
 S_{\ln}=\frac{A}{4l_{p}^{2}}+ \frac{\widetilde{\alpha}}{4}\ln
 \left(\frac{A}{l_{p}^{2}}\right),
\end{equation}
omitting the Boltzmann constant $k_{B}$ as a factor and assuming
it to be equal to unity in what follows. Higher-order corrections
may be derived using the Taylor-series expansion in terms of the
small parameter $l_{p}^{2}/A$
\begin{equation}\label{GUP5.2}
 S_{GUP}=\frac{A}{4l_{p}^{2}}+ \frac{\widetilde{\alpha}}{4}\ln
 \left(\frac{A}{l_{p}^{2}}\right)+\sum_{n=1}^{\infty}\widetilde{c}_{n}
\left({A\over l_p^2} \right)^{-n}+ \rm{const}
\end{equation}
in a similar way to the Taylor-series expansion of the right-hand
side in (\ref{GUP5})
 in terms of the small parameter $4l_{p}^{2}/A$. This is valid as GUP gives
the ultraviolet cutoff at the level of $l_{min}\sim l_{p}$.
\\In this way at high energies we have $S\rightarrow S_{GUP}$  and hence
$N\rightarrow N_{GUP}$. Assuming in the notation of
\cite{Verlinde} that
\begin{equation}\label{Verl2}
 S=\frac{1}{4}N,
\end{equation}
we directly obtain
\begin{equation}\label{GUP5}
N_{GUP}= \frac{A}{l_{p}^{2}}+ \widetilde{\alpha}\ln
 \left(\frac{A}{l_{p}^{2}}\right)+4\sum_{n=1}^{\infty}\widetilde{c}_{n}
\left({A\over l_p^2} \right)^{-n}+ \rm{const},.
\end{equation}
Now, coming back to \cite{Verlinde}, in terms of $N_{GUP}$ we can
define the holographic screen area, as measured at high energies,
by
\begin{equation}\label{GUP6}
A_{GUP} \equiv G\hbar  N_{GUP},
\end{equation}
where $G$  and  $\hbar$ -- gravitational and Planck constants,
respectively, and $N_{GUP}$  is given by (\ref{GUP5}). Considering
that we, similar to \cite{Verlinde},
 assume that the speed of light $c=1$, then, according to $l_p^2=G\hbar$, from
(\ref{GUP5}) and (\ref{GUP6}) we have
\begin{equation}\label{GUP7}
A_{GUP} =A+G\hbar \widetilde{\alpha}\ln
\left(\frac{A}{G\hbar}\right) +4G\hbar
\sum_{n=1}^{\infty}\widetilde{c}_{n} \left({A\over G\hbar}
\right)^{-n}+ \rm{const}\;.
\end{equation}
In this case an exact value of the constant in the right-hand side
of (\ref{GUP7}) is of no great importance as further we need the
relation (\ref{Verl1}), being primarily interested in $dA_{GUP}$
rather than in $A_{GUP}$, i.e. the constant in the right-hand side
of (\ref{GUP7}) is insignificant. So, (\ref{Verl1}) has a fairly
definite analog at high energies
\begin{equation}\label{Verl2}
 dN_{GUP}={dA_{GUP}\over G\hbar}
\end{equation}
that on going to the known low energies gives (\ref{Verl1}). There
is a single considerable difference, in \cite{Verlinde} the
quantity $N$ was defined in terms of $A$ and $dN$ was defined in
terms of $dA$ but in the case under study the situation is
opposite: $A_{GUP}$  is defined in terms of $N_{GUP}$ and
$dA_{GUP}$ in terms of $dN_{GUP}$. The logic series is here as
follows:
\begin{equation}\label{Verl3}
  A\Rightarrow N \Rightarrow
N_{GUP}\Rightarrow A_{GUP}.
\end{equation}
The GUP-correction problem of the temperature $T$ for the
arbitrary holographic screen $\cal S$ has been studied in
\cite{Setare1}. Actually, this case is identical to the case of
the Schwarzschild black hole and $T_{GUP}$ was derived as a series
(formulas (1),(7) from \cite{Setare1})
\begin{equation}\label{Setare}
T_{GUP}=T(1+\Theta_{T} T^2+...)=T+ \Theta_{T}
T^3+...=T+\widetilde{T}_{GUP},
\end{equation}
where the factors in the right-hand side(\ref{Setare}) may be computed in the explicit form and at low energies
$\widetilde{T}_{GUP}\rightarrow 0$.
\\Thus, we can have a GUP -
analog of Komar's mass in ((5.33) from \cite{Verlinde})
\begin{equation}\label{Verl10}
  M_{GUP}={1\over 2} \int_{\cal
S}T_{GUP}dN_{GUP}={1\over 2} \int_{\cal
S}(T+\widetilde{T}_{GUP})dN_{GUP}={1\over 2 G\hbar} \int_{\cal
S}T_{GUP}dA_{GUP},
\end{equation}
 that in the low-energy limit gives the well-known Komar
 formula \cite{Komar},(\cite{Wald}, p.289).
\\It is clear that the   "GUP-deformed Komar's mass"   $M_{GUP}$ in the first term (\ref{Verl10})
as a summand has the known Komar's mass \cite{Komar},((11.2.9 -
11.2.10),\cite{Wald}) ((5.34), \cite{Verlinde})
\begin{equation}
\label{Komar1}
 M ={1\over 4\pi G}\int_{\cal S}T dA.
\end{equation}

\section{$N_{GUP}$,$A_{GUP}$ and $M_{GUP}$ in Terms of  Unified Small Parameter}
If feasible, it is desirable to express all the above-derived
fundamental quantities in terms of a unified parameter. As shown
by the author in \cite{shalyt-entropy2}, \cite{shalyt-IJMPD}, this
is possible for black holes within the scope of GUP and a role of
the unified small parameter is played by   the parameter
introduced previously in \cite{shalyt1}--\cite{shalyt10} as
follows:
\begin{equation}\label{D1}
\alpha_{x}=l_{min}^{2}/x^{2},
\end{equation}
where $x$ is the measuring scale, $l_{min}\sim l_{p}$  by virtue of GUP
(\ref{GUP1}), and $0<\alpha\leq 1/4$.
\\ Obviously, the principal results obtained in \cite{shalyt-entropy2}, \cite{shalyt-IJMPD}
remain in force for an arbitrary screen ${\cal S}$ and may be
applied to the quantities $N_{GUP}$,$A_{GUP}$, $M_{GUP}$ defined
in the preceding Section.
\\ Substituting from GUP (\ref{GUP1}) $l_{min}=2\surd\alpha^{\prime}
l_{p}$  and using the formula $A=4\pi R^{2}$, where $R$ is the
radius of the screen ${\cal S}$, we get  $N_{GUP}$ (\ref{GUP5}) of
the following form:
\begin{equation}\label{GT6.3}
N_{GUP}=
 N+\widetilde{\alpha}\ln (\sigma \alpha^{-1}_{R})
+4\sum_{n=1}^{\infty}\widetilde{c}_{n}\sigma^{-n} \alpha^{n}_{R}+
\rm{const}.
\end{equation}
Here $\alpha_{R}$ is a value of $\alpha$ parameter at the point
$R$ and $\sigma$ is a equal  to $16\alpha^{\prime}\pi$\\. It is
convenient to refer to the form $N_{GUP}$ derived in (\ref{GT6.3})
as to the $\alpha$ --representation.
\\Using (\ref{GUP6}) and (\ref{GUP7}),
 we can easily obtain $\alpha$ -- representation
for $A_{GUP}$
\begin{equation}\label{GT6.3.2}
A_{GUP}=A+\widetilde{\alpha}G\hbar\ln (\sigma \alpha^{-1}_{R})
+4G\hbar\sum_{n=1}^{\infty}\widetilde{c}_{n}\sigma^{-n}
\alpha^{n}_{R}+ \rm{const}.
\end{equation}
Also, it is clear that $M_{GUP}$ (\ref{Verl10})
 may be derived in terms of $\alpha_{R}$.
\\Here $\alpha_{x}$  is considered
 as a deformation parameter for the Heisenberg algebra  on going from HUP   to GUP.
 Generally speaking, initially the construction of such a deformation
 was realized with other parameters
(e.g. \cite{Magg1},\cite{Kempf}). But it is easily shown that QFT
parameter of the deformations associated with GUP may be expressed
in terms of the parameter $\alpha$ that has been introduced in the
approach to the density matrix deformation
\cite{shalyt-aip},\cite{shalyt-entropy2}. Here the notation of
\cite {Kim2} is used. Then (\cite{shalyt-entropy2}, p. 943)
\begin{equation} \label{comm1}
[\vec{x}, \vec{p}]=i\hbar(1+\beta^2\vec{p}^2+...)
\end{equation}
and
\begin{equation}\label{comm2}
\Delta x_{\rm min}\approx \hbar\sqrt{\beta}\sim l_{p}.
\end{equation}
In this case from (\ref{comm1}),(\ref{comm2}) it follows that $\beta\sim
{\bf 1/p^{2}}$, and for $x_{\rm min}\sim l_{p}$, $\beta$ corresponding to $x_{\rm min}$ is nothing else but
\begin{equation}\label{comm3}
\beta\sim  1/P_{pl}^{2},
\end{equation}
where $P_{pl}$ is Planck's momentum: $P_{pl}= \hbar/l_{p}$.
\\In this manner $\beta$ is varying over the following interval:
\begin{equation}\label{comm4}
\lambda/P_{pl}^{2}\leq \beta,
\end{equation}
where $\lambda$  is  a numerical factor  and the second term in
(\ref{comm1}) is accurately reproduced in the momentum representation
(up to the numerical factor) by $\alpha_{x}=l^{2}_{min}/x^{2}\sim
l^{2}_{p}/x^{2}=p^{2}/P_{pl}^{2}$
\begin{equation} \label{comm5}
[\vec{x},\vec{p}]=i\hbar(1+\beta^2\vec{p}^2+...)=i\hbar(1+a_{1}\alpha_{x}+a_{2}\alpha_{x}^{2}+...).
\end{equation}
In the case under study convenience of using $\alpha_{x}$ stems
from its smallness, its dimensionless character, and ability to
test changes in the radius $R$ of the holographic screen ${\cal
S}$.
\section{Quantum Corrections to the Principal Result and Ultraviolet Limit}
Based on the aforesaid, we can proceed to generalization of the
results from Section   5.2 of \cite{Verlinde} and to derivation of
equations for a gravitational field within the scope of GUP.
\\ We must consider two absolutely different cases.
\\
\subsection{Quantum Corrections to the Principal Result}
It is assumed that the screen radius is given by
${\cal S}$
\begin{equation} \label{R1.1}
 R \gg  l_{p}.
\end{equation}
In terms of the deformation parameter $\alpha_{x}$ introduced in
the previous Section, we have
\begin{equation}\label{R2}
\alpha_{R}\ll 1/4.
\end{equation}
So far we are not concerned with redefinition of the lower limit
for $\alpha_{R}$. This is interesting when going to the infrared
limit.
\\ Then the principal result from the final part of Section 5.2 in
\cite{Verlinde} remains valid owing to the replacement of $M$
(formula (5.33) from \cite{Verlinde})  by  $M_{GUP}=
M_{GUP}[\alpha_{R}]$ (\ref{Verl10}).  The "$\alpha_{R}$ --
complement" (i.e. the difference
$\widetilde{M}[\alpha_{R}]=M_{GUP}[\alpha_{R}]-M$) to $M$ will be
simply a (small) quantum correction for the principal result.
\\In this way, because in the case of GUP-corrections the left side of
formula (5.33) and hence the left side in
formulas (5.34), (5.35) from \cite{Verlinde} is dependent on
$\alpha_{R}$, the right sides of the corresponding formulas are also dependent on $\alpha_{R}$,in particular  the quantities
$T_{ab},R_{ab},g_{ab}$ from (5.37).
\\ But in fact this relation at low energies ((\ref{R1.1}) or
(\ref{R2})) is insignificant since $\alpha_{R}$ at low energies
(as distinct from high Planck’s energies)is varying very slowly, practically showing continuity though being discrete in character.
\\ Indeed, as long as there is the minimal length $l_{min}\sim
l_{p}$, all the lengths measured are its multiples and hence $\alpha_{R}$  is a discrete non-uniformly varying quantity. Then, due to ((\ref{R1.1})
or (\ref{R2})), the difference between two   successive values of $\alpha_{R}$  is as follows:
\begin{equation}\label{R2.1}
\Delta_{min}[\alpha_{R}]=\alpha_{R}-\alpha_{R+l_{min}}\sim
\frac{l^{3}_{min}}{R^{3}},
\end{equation}
for $R \gg  l_{p}$  or, that is the same, for $R \gg  l_{min}$
giving a value close to zero.
\\ And assuming in this case that $\alpha_{R}$
is continuously varying from $R$ and all the quantities in Section
5.2 of \cite{Verlinde} are also continuously dependent on
$\alpha_{R}$ (\ref{R2}), we can write down the ("$\alpha$ --
analog" of formula (5.37) in \cite{Verlinde})as
\begin{equation}\label{R3}
 2\int_{\Sigma}\left(T_{ab}[\alpha] -{1\over
2}T[\alpha]g_{ab}[\alpha]\right) n^a \xi^b dV= {1\over 4\pi
G}\int_{\Sigma}R_{ab}[\alpha] n^a \xi^b dV,
\end{equation}
 where the dependence of $T_{ab}[\alpha]$  and  $R_{ab}[\alpha]$  on
 $\alpha=\alpha_{R}$  is completely determined, in accordance with \cite{Wald},\cite{Verlinde},
 by the integral  $ M_{GUP}[\alpha]$ (\ref{Verl10}).
 \\ Besides, it is assumed that $n^a $ and $\xi^b$ are dependent on
 $\alpha$, though the dependence is dropped.
 \\Next, similar to \cite{Verlinde},
 from (\ref{R3})we can derive the  {\bf $\alpha$-deformed} Einstein
 Equations using the method from \cite{Jac1}. Note that both
 this method and its minor modification given in
(\cite{Verlinde}, end of Section 5.2) in this case are valid   because $\alpha_{R}$
 is small and continuous, the whole system being continuously dependent on it.
\\ Solutions of the {\bf $\alpha$-deformed} Einstein
 Equations represent a series in $\alpha_{R}$, and
  for $\alpha_{R}\rightarrow 0$  or  for $\alpha^{\prime}=0$ from
 formula (\ref{GUP1}) become the corresponding solutions of (Section 5.2 in
\cite{Verlinde}).
\\Using the result obtained in \cite{Magnon},
 we can easily extend the above result to the case with a nonzero cosmological term
$\Lambda\neq 0$. In \cite{Magnon} Komar's formula was generalized
to the case of a nonzero $\Lambda$. All the arguments from
(Section 5.2 of \cite{Verlinde}) in this case remain valid and
formula (5.37) takes the following form:
\begin{equation}\label{Einstein-new}
2\int_{\Sigma}\left(T_{ab} -{1\over 2}Tg_{ab}\right) n^a \xi^b dV=
{1\over 4\pi G}\int_{\Sigma}(R_{ab}+\Lambda g_{ab}) n^a \xi^b dV.
\end{equation}
 We can easily obtain the $\alpha$ - analog of the last formula
with the dynamic cosmological term $\Lambda(\alpha)$  as a
corresponding complement to the right-hand side (\ref{R3}).
Analysis of the relationship between $\Lambda$ and $\alpha$,
applicable in this case as well, will be given in Section 4.2.
\\
\subsection{Ultraviolet Limit}
In the case in question we suggest that the screen ${\cal S}$ has
a radius on the order of several Planck’s lengths
\begin{equation}\label{L1.1}
 R \approx  \xi l_{min}=2 \alpha^{\prime}\xi  l_{p},
\end{equation}
where $\xi$ is a  number on the order of  1 or
\begin{equation}\label{L2}
\alpha_{R} \approx 1/4.
\end{equation}
The problem is which object
puts the limit for such a screen ${\cal S}$. It may be assumed
that if $ T_{ab}\neq 0$ then the object may be represented only by
Planck’s black hole or by a micro-black hole with a radius on the
order of several Planck’s lengths.
\\Clearly, the methods of \cite{Verlinde} and
\cite{Jac1} are not in force for such screen ${\cal S}$.
Specifically, there are no classical analogs of $N$, $T$, and $M$
for the screen. Moreover, it is impossible to use the result of
\cite{Jac1} as "a very small region the space-time" is no longer
"an approximate Minkowski space"  \cite{Verlinde}.
\\Also, such micro-black hole is a horizon space, jet at high energies (Planck scales).
As is known, for horizon spaces, black holes in particular, at low
energies (semiclassical approximation) the results of
\cite{Padm13} are valid.
 \\{\bf At the horizon (and we are interested in this case only)
 Einstein's field Equations may be written as a thermodynamic identity}
(\cite{Padm13} formula (119))
\begin{equation}\label{GT12}
   \underbrace{\frac{{{\hbar}} cf'(a)}{4\pi}}_{\displaystyle{k_BT}}
    \ \underbrace{\frac{c^3}{G{{\hbar}}}d\left( \frac{1}{ 4} 4\pi a^2 \right)}_{
    \displaystyle{dS}}
  \ \underbrace{-\ \frac{1}{2}\frac{c^4 da}{G}}_{
    \displaystyle{-dE}}
 = \underbrace{P d \left( \frac{4\pi}{ 3}  a^3 \right)  }_{
    \displaystyle{P\, dV}}
\end{equation}
where $R=a$ is  a radius of a black hole (i.e. of the screen
${\cal S}$), $P =T^{R}_{R}$ is the trace of the momentum-energy
tensor and radial pressure, and the horizon location will be given
by simple zero of the function $f(R)$, at $R=a$.
\\The main ingredients of (\ref{GT12}) may be written in terms of the deformation
parameter $\alpha$ with the coefficients containing only the
numerical factors and fundamental constants \cite{shalyt-IJMPD}.
\\Also, the work \cite{shalyt-IJMPD} presents two possible variants of high-energy (Planck) $\alpha$
-deformation $\alpha \rightarrow 1/4$ (\ref{GT12}).
\\Hereinafter, we assume that the energy – momentum tensor of matter fields is not traceless
\begin{equation}\label{GT12.1}
 T^{a}_{a}\neq 0,
\end{equation}
similar, in particular, to the case under study
(\ref{GT12}) $P
=T^{R}_{R}\neq 0$
\\
\\{\bf 1. Case of equilibrium thermodynamics (\cite{shalyt-IJMPD}, section
(6.1))}
\\
\\ In this case it is assumed that in the high-energy
(ultraviolet (UV))limit the thermodynamic identity (\ref{GT12})
 is retained but now all the quantities involved
in this identity become $\alpha$-deformed ($\alpha \rightarrow
1/4$). All the quantities $\Upsilon$ in (\ref{GT12}) are replaced
by the corresponding quantities $\Upsilon_{GUP}$ with the
subscript GUP. Then the high-energy $\alpha$-deformation of
equation (\ref{GT12}) takes the form
\begin{equation}\label{GT8.GUP}
k_{B}T_{GUP}(\alpha)dS_{GUP}(\alpha)-dE_{GUP}(\alpha)=P(\alpha)dV_{GUP}(\alpha).
\end{equation}
Substituting into (\ref{GT8.GUP}) the corresponding quantities
\\$T_{GUP}(\alpha),S_{GUP}(\alpha),E_{GUP}(\alpha),V_{GUP}(\alpha),P(\alpha)$
and expanding them into a Laurent series in terms of $\alpha$,
close to high values of $\alpha$, specifically close to
$\alpha=1/4$, we can derive a solution for the high energy
$\alpha$-deformation of the general relativity (\ref{GT8.GUP}) as a
function of $P(\alpha)$. Provided at high
energies the generalization of (\ref{GT12}) to (\ref{GT8.GUP})is
possible, we can have the high-energy $\alpha$-deformation of the
metric.
\\ It is noteworthy that in (\ref{GT8.GUP}) $T_{GUP}$
this time is calculated from (\cite{Park}, formula (10))
 \begin{eqnarray}\label{Verl7.micro}
T^{BH}_{GUP}=\frac{1}{4\pi}\frac{\hbar R}{2 \alpha^{\prime 2}
l_p^2}[1- \sqrt{1-\frac{\alpha^{\prime 2} l_p^2}{R^2}}]=\nonumber
\\ \frac{\hbar \alpha^{-1}_{R} }{4 \pi
\alpha^{\prime}l_{p}}[1-(1-\alpha_{R})^{1/2}]
\end{eqnarray}
with subsequent replacement of $l_{p}$ by $\sqrt{G\hbar}$ for
$c=1$.
\\It is especially interesting to consider the following case.
\\
\\{\bf 1. Case of nonequilibrium thermodynamics (\cite{shalyt-IJMPD}, section
(6.2))}
\\
\\ In this case the $\alpha$ - dependent dynamic cosmological term $\Lambda(\alpha)\neq 0$
appears in the right-hand side of(\ref{GT8.GUP}).
Then, with the addition of
$\Lambda(\alpha)\neq 0$ , the $\alpha$ -- representation
(\ref{GT8.GUP})(for $\hbar=1$) is given as follows((\cite{shalyt-IJMPD}, formula (53)):
 \begin{equation}\label{GT16.B}
 -\alpha^{2}f'(\alpha)-\frac{1}{2}\alpha=16\pi \alpha^{\prime 2}P(\alpha)G^{2}-G\Lambda(\alpha),
 \end{equation}
 where $\alpha=\alpha_{R}\approx 1/4$,
 \begin{equation}\label{GT8.GUP2}
f'(\alpha)=4\pi k_{B}T_{GUP}(\alpha)
\end{equation}
and the derivative in the left-hand side of (\ref{GT8.GUP2}) is taken with respect to  $\alpha$.
\\$\Lambda(\alpha)$ in the right-hand side of (\ref{GT8.GUP2})
 may be subjected to a series expansion in terms of $\alpha$,
in compliance with the holographic principle
\cite{Hooft1}--\cite{Bou3} as applied to the Universe
\cite{Sussk1}. In \cite{Bal},\cite{shalyt-aip},
\cite{shalyt-entropy2},\cite{shalyt-IJMPD}  in the leading order
this expansion results in the first power, i.e. we have
\begin{equation}\label{DE7}
\Lambda(\alpha_{R}) \sim \alpha_{R}\Lambda_{p},
\end{equation}
where $\Lambda_{p}$ -- initial value of
$\Lambda\approx\Lambda_{1/4}$ derived using the well-known
procedure of "summation over all zero modes " and the  Planck
momentum cutoff \cite{Zel1},\cite{Wein1}. Actually, (\ref{DE7})is
in a good agreement with the observable
$\Lambda=\Lambda_{observ}$. Because a radius of the visible part
of the Universe is given as $R=R_{Univ}\approx 10^{28} cm$, it is
clear that $\alpha_{R}\approx 10^{-122}$  and (\ref{DE7})is
completely consistent with the experiment  \cite{Wein1}.
\\Note that, proceeding directly from a quantum field theory but without
the use of the holographic principle, we can have only a rough
estimate of $\Lambda$ that, on the whole, is at variance with
$\Lambda_{observ}$. Such an estimate may be obtained in different
ways:  by simulation \cite{Mark1}; using the cutoff \cite{Zel1}
but now in the infrared limit; with the use of the Generalized
Uncertainty Principle for the pair $(\Lambda, V)$, where $V$ --
four-dimensional volume \cite{shalyt-aip}, \cite{shalyt-entropy2}.
In the $\alpha$--representation in this case the expansion in
terms of $\alpha$ results in the second leading order
\begin{equation}\label{DE7.1}
\Lambda(\alpha_{R}) \sim \alpha^{2}_{R}\Lambda_{p},
\end{equation}
that, obviously, is at variance with the accepted facts. \\
\section{Conclusion}
I. A very interesting case of the zero energy-momentum tensor for
matter fields $ T_{ab}=0$ and , specifically the case of
$P(\alpha)=0$  in the right-hand side of (\ref{GT16.B}) has
remained beyond the scope of the final Section. We can state the
problem more specifically: for which conditions in this case we
can derive a solution in the form of the de Sitter space with
large values of $\alpha$?
 \\
 This problem is also important when we try to find whether it
 is possible to derive the initial inflation conditions \cite{Kolb},\cite{Baum}
 for $ T_{ab}=0$ on the basis of the foregoing analysis.
 \\
 \\ Note that the dynamic cosmological term
 $\Lambda(\alpha)$ correlates well with inflation models
\cite{Kolb},\cite{Baum} as the latter require a very high
$\Lambda$ at the early stages of the Universe, and this is
distinct from $\Lambda=\Lambda_{exper}$  in the modern period.
 Of great interest is the recent work \cite{Polyak}, where a mechanism
of the vacuum energy decay in the de Sitter space is established
to support a dynamic nature of $\Lambda$.
\\
 II. The deformation parameter $\alpha_{R}$  has a double meaning.
\\ As $\kappa = 1/R$ -- curvature with the radius $R$,
 $\alpha_{R}=\kappa^{2}l^{2}_{min}$ is nothing else but the squared
 curvature multiplied by the squared minimal area and is  explicitly dependent on the energy $E$.
 On the other hand, it is seen that, at least at the known energies, from the definition of the bit number $N$ in
(\cite{Verlinde}, formula (3.10)) we get $\alpha_{R}\sim
   1/N$. But, because $\alpha_{R}=\alpha_{R}[E]$, this suggests that $N=N[E]$,
 as demonstrated in the text on going to higher energies
\begin{equation}\label{Concll}
 N \Rightarrow
N_{GUP}.
\end{equation}
 Nevertheless, on going to lower energies,
i.e. in the infrared limit, the same should be true: the bit
number must be a function of energy.
\\This problem and the relevant questions touched upon in this paper
 will be further considered in subsequent works of the author.

\end{document}